\documentclass[aps,prb,twocolumn,reprint,superscriptaddress,floatfix]{revtex4}
\usepackage{graphicx}
\usepackage{amsmath}
\usepackage{color}
\usepackage{placeins}
\usepackage[hidelinks]{hyperref}

\begin{document}

\title{The Rise, Fall and Rebirth of an Oxide: growth and evolution of Bismuth Oxide on Au(111)}

\author{Alberto Turoldo}
\affiliation{Department of Physics, University of Trieste, Trieste, Italy}

\author{Marco Bianchi}
\thanks{Corresponding author}
\email{marco.bianchi@elettra.eu}
\affiliation{Elettra - Sincrotrone Trieste S.C.p.A, Trieste, Italy}

\author{Deborah Perco}
\affiliation{Department of Physics, University of Trieste, Trieste, Italy}

\author{Andrea Berti}
\affiliation{Department of Physics, University of Trieste, Trieste, Italy}

\author{Philip Hofmann}
\affiliation{Department of Physics and Astronomy, Aarhus University, Aarhus, Denmark}

\author{Alessandro Baraldi}
\affiliation{Elettra - Sincrotrone Trieste S.C.p.A, Trieste, Italy}
\affiliation{Department of Physics, University of Trieste, Trieste, Italy}

\author{Silvano Lizzit}
\affiliation{Elettra - Sincrotrone Trieste S.C.p.A, Trieste, Italy}


\begin{abstract}
We report a comprehensive, multi-technique study of bismuth oxide growth on Au(111) under different conditions such as temperature and exposure to molecular oxygen. By combining synchrotron-based X-ray photoelectron spectroscopy and diffraction with low-energy electron diffraction and scanning tunneling microscopy, we elucidate the structural evolution of the system during controlled oxidation and subsequent annealing. We find that Bi deposition induces well-defined surface reconstructions, whereas oxidation triggers the formation of a complex sequence of bismuth oxide domains. High-resolution spectroscopic and diffraction data enable us to pinpoint the structure of the oxide consistent with the $(201)$ surface of $\beta$-Bi$_2$O$_3$. In addition, angle resolved photoemission experiments and work function measurements reveal substantial electronic modifications triggered by a complex interplay of segregation of Bi and Au. Notably, a work function of 3.4~eV is achieved for oxidation at 423~K. These results provide benchmark structural and electronic insights into the Bi oxide/Au(111) system and establish a framework for integrating Bi$_2$O$_3$ as a contact material in devices using two-dimensional semiconductors, where it can enable low contact resistance.
\end{abstract}

\keywords{Bismuth Oxide, Two-dimensional materials, Contact engineering, Charge-transfer doping}

\maketitle

\section{Introduction}\label{sec1}

The study of semimetal-based interfaces for next-generation electronics requiring ultralow contact resistance has attracted significant attention, since recent reports indicate that Bi- or Sb-based contacts can suppress metal-induced gap states at transition metal dichalcogenides (TMDs) \cite{Shen2021,Li2023,Su2023}.
Similarly, the ability to tune the electronic properties of two-dimensional (2D) systems, such as achieving degenerate n-doping, has been observed when contacting thin oxide layers  with TMDs \cite{Rai:2015aa,Leonhardt:2019aa,McClellan:2021aa}, demonstrating significant upside as a versatile platform for integration also with 2D dielectric nanosheets and other 2D materials such as graphene. A notable example is the formation of graphene–Al$_2$O$_3$ nanosheet interfaces, where an ultrathin alumina layer enables electronic decoupling while preserving structural integrity \cite{Omiciuolo:2014aa}. 
The great technical potential of these approaches is limited by the current understanding of the dominant mechanisms governing the desired effects, which is largely due to the challenges associated with the characterization not only of the interface, but more fundamentally of the oxides, which are inherently prone to defects, vacancies, and other structural and stoichiometric irregularities.
This is the case of Bismuth, for which several oxides with distinct stoichiometries exist, among which Bi$_2$O$_3$ is the most stable and widely used in technological applications. 
Bi$_2$O$_3$ is itself a polymorphic material of considerable technological relevance, exhibiting a wide range of functional properties that have been extensively exploited over the past decades. 
Its applications extend from resistive components \cite{Maeder2013} and heterogeneous catalysis  \cite{Lou2014,Chen2021} to visible-light photocatalysis \cite{Xiao2013,Tian2021,Hashemi2025} and, more recently, to p-type semiconducting channels for nanoscale transistors  \cite{Xiong2025}. 
Despite this broad technological interest and the body of literature about it \cite{Guertler1903, Gattow1964, Rao1969, Harwig1979, Gualtieri1997, Cornei2006, Drache2007}, a detailed understanding of the growth and oxidation of Bi thin films under well-controlled ultra-high vacuum (UHV) conditions remains scarce and incomplete. This gap underscores the need for a systematic characterization of the structural and electronic properties of Bi and its oxides at metal surfaces like Au(111)—which is widely used as  a supporting substrate also for 2D materials.
In this work, we address this gap through a multi-technique study of Bi and Bi oxide growth on Au(111), combining synchrotron-based structural and spectroscopic probes with diffraction and microscopy techniques to resolve the evolution of the interface under controlled preparation conditions.

\begin{figure*}[htbp]
	\includegraphics[width=0.8\textwidth]{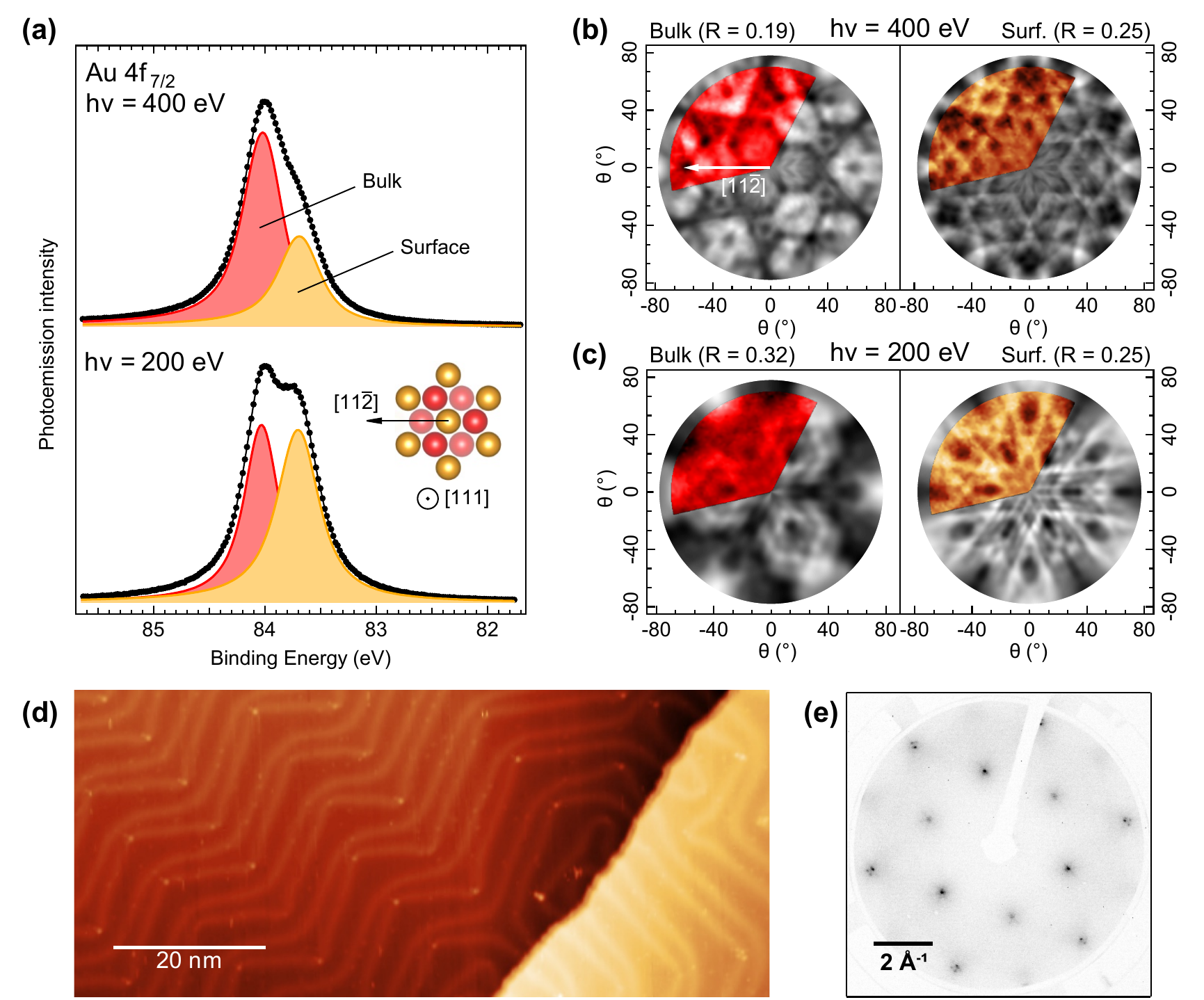}
	\caption{Characterization of the Au(111) surface. (a) Au 4$f_{7/2}$ core level spectrum acquired at photon energies $h\nu = 400$ and $200$~eV. Acquired data are displayed as points while fitted bulk and surface components in solid areas. The atomic model of the surface is presented. (b)-(c) Stereographic projection of the modulation function $\chi$ for the bulk and surface components acquired at $h\nu = 400$~eV and $h\nu = 200$~eV (corresponding to a kinetic energy of $316$ and $116$~eV, respectively). Acquired data (colored section) are compared with XPD simulation (greyscale) and the $[11\bar{2}]$ direction are indicated as reference. The reported R-factors, for this case only, are calculated after matching the modulation function amplitude of the simulated data to the amplitude of the experimental data. (d) STM image displaying the herringbone reconstruction of the Au(111) surface ($I = 0.40$~nA, $V = 1.60$~V), with low coverage of Bi visible as faint bright protrusions on the herringbone elbows. (e) LEED pattern obtained at $E_k = 137$~eV, the herringbone reconstruction generates the additional spots surrounding the first order Au(111) diffraction peaks.}
	\label{fig:au111}
\end{figure*}

\section{Experimental Methods}\label{sec4}

Experiments were carried out at the SuperESCA beamline of Elettra Sincrotrone Trieste \cite{Abrami1995} (IT), including its CoSMoS branch line equipped with a monochromatized Al K$_\alpha$ X-ray source, and at the SGM3 beamline of ASTRID2 (DK) \cite{Hoffmann:2004aa}. 

Surface characterization combined complementary imaging, diffraction and spectroscopic techniques. X-ray Photoelectron Spectroscopy (XPS) was employed in all laboratories as a key tool to determine chemical composition and oxidation states\cite{Bignardi:2023}, while X-ray Photoelectron Diffraction (XPD) performed on the SuperESCA beamline provided local structural information. Angle-Resolved Photoemission Spectroscopy (ARPES) was used to probe the electronic band structure at the SGM3 beamline. 
Long-range surface order was assessed by Low Energy Electron Diffraction (LEED) at all the facilities, and real-space morphology was investigated by Scanning Tunneling Microscopy (STM) on CoSMoS and SGM3.

\begin{figure*}
	\includegraphics[width=0.8\textwidth]{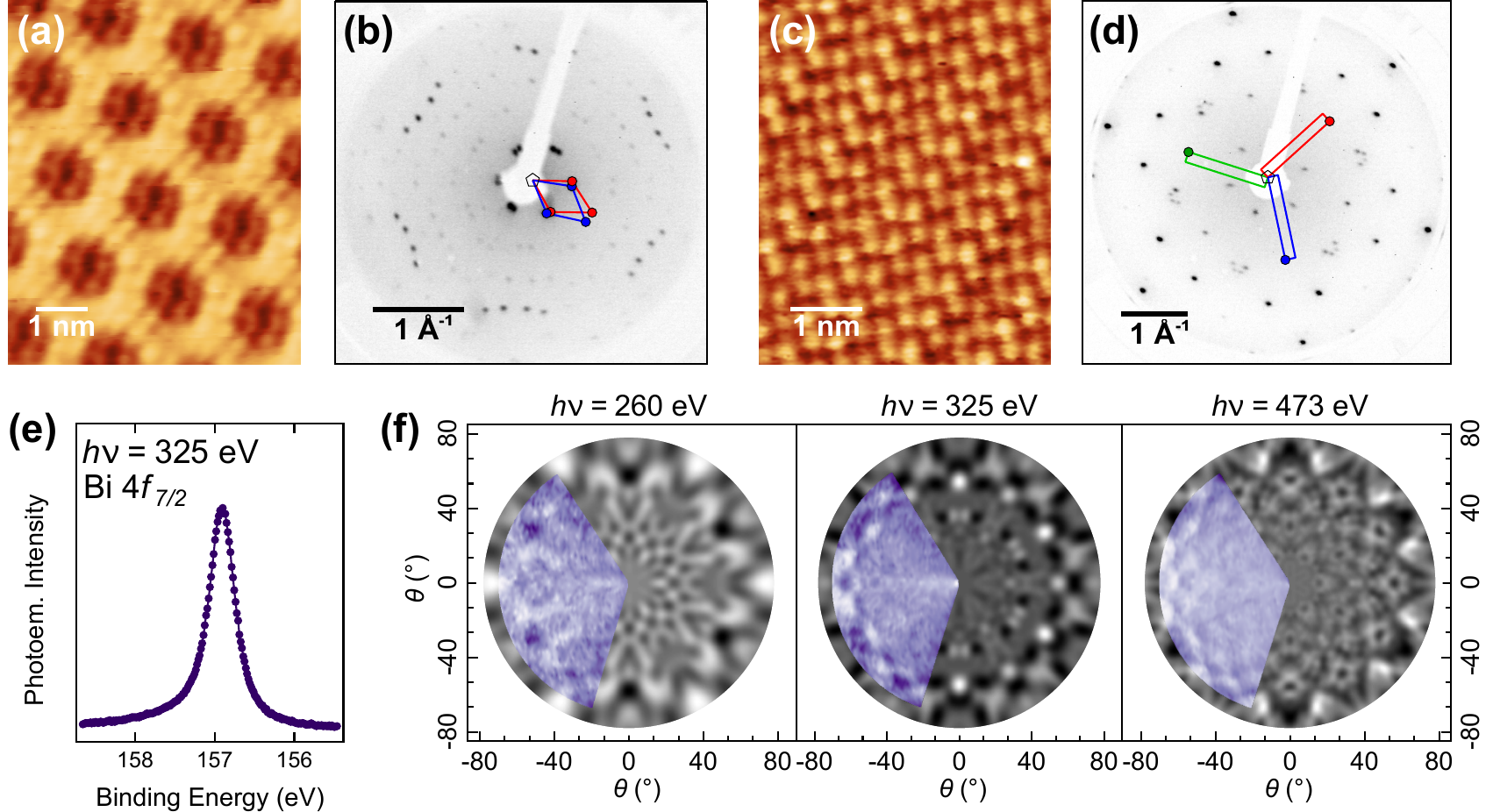}
	\caption{Phases for Bi deposition on Au(111). (a)-(b) $\left(\sqrt{37}\times\sqrt{37}\right)R25.3^\circ$ phase. (a) STM image displaying honeycomb structure ($I = 0.22$~nA, $V = 0.23$~V); (b) LEED image acquired at $E_\mathrm{kin} = 38$~eV with highlighted unit cells (red/blue) associated to different rotational domains. (c)-(d) $(P \times \sqrt{3})$ phase, with $P = 14$. (c) STM image showing a striped pattern associated to the moiré supercells ($I = 0.12$~nA, $V = 1.95$~V).  Phases with lower P were also observed that are not presented here as out of the scope of this paper. (d) LEED image acquired at $E_\mathrm{kin} = 38$~eV with highlighted unit cells (red/green/blue) associated to different rotational domains. (e) High resolution Bi 4$f_{7/2}$ core level spectrum obtained after bismuth deposition in the multilayer regime, acquired at $h\nu = 325$~eV. (f) Stereographic projections of the modulation function $\chi$ of the Bi 4$f_{7/2}$ peak at different kinetic energies, same multilayer conditions as in (e); The acquired data (colored section) are compared with XPD simulations (greyscale).}
	\label{fig:bi_phases}
\end{figure*}
Clean Au(111) surfaces were prepared by repeated cycles of Ar$^+$ sputtering and subsequent annealing up to 970~K. Surface cleanliness and long-range order were verified by XPS, LEED, and STM. Bismuth was deposited onto Au(111) under ultra-high vacuum (UHV) conditions from a home-built evaporator (Tantalum boat thermal evaporator), with the substrate maintained at approximately 373~K. The temperature was monitored using a K-type thermocouple in direct contact with the crystal. The chemical purity of the deposited Bi and the absence of contaminants were confirmed by XPS. 

Oxidation was performed by exposing the Bi-covered surface to high-purity O$_2$ through a home-made doser positioned close to the sample, enabling a local pressure up to $10^{-3}$~mbar while maintaining a chamber pressure of $10^{-5}$~mbar during dose \cite{Lizzit:2012aa}. Oxidation at higher pressures was performed in a dedicated gas-cell connected to the UHV chambers.

XPS measurements were performed at room temperature in normal emission geometry (unless otherwise specified) using synchrotron radiation and a Phoibos 150 analyzer, achieving a combined energy resolution better than 80~meV.
 All the spectra are shown after subtraction of a linear background and calibration of the binding energy ($E_b$) with respect to the measured Fermi level ($E_{F}$). The work function $\Phi$ was determined from the kinetic energies, obtained by photoemission measurements, where the Fermi energy and the secondary electron cutoff appear in the spectrum\cite{Baddorf2023} ($E_{F}^{\mathrm{kin}}$ and $E_\mathrm{cut}^{\mathrm{kin}}$ respectively) according to
\begin{eqnarray}
\Phi = h\nu - \left( E_{F}^{\mathrm{kin}} - E_\mathrm{cut}^{\mathrm{kin}} \right).
\end{eqnarray} 
A  bias voltage of $-7$~V was applied to the sample during these measurements. All the spectra acquired for the secondary electron cutoff identification employed photon energies of 109.95~eV or 145.05~eV.

XPD data are presented as stereographic projections of the modulation function $\chi$, defined as in Ref.~ \cite{Orlando2014}. Experimental patterns ($\chi_{exp}$) were compared with multiple-scattering simulations ($\chi_{sim}$) based on trial atomic configurations generated using the Electron Diffraction in Atomic Clusters (EDAC) package  \cite{GarciadeAbajo2001}.
Starting crystallographic parameters were obtained from literature\cite{Maeland:1964,Cucka:1962,PerezMezcua:2016aa} using the Materials Cloud \cite{Talirz2020}. 
 Agreement between simulated and experimental patterns is quantified in terms of the R-factor\cite{Woodruff:2007} defined as 
\begin{eqnarray}
R=\frac{\sum _i (\chi_{i, exp}-\chi_{i,sim})^2}{\sum _i (\chi_{i, exp}^2+\chi_{i,sim}^2)}.
\end{eqnarray}

STM topographs were acquired at room temperature (RT) using  Aarhus STMs equipped with a PtIr tip. Image processing and analysis were performed with the WSxM software package  \cite{Horcas2007} while structural models were visualized using the VESTA software  \cite{Momma2011}. LEED kinematic simulations were performed with ProLEED Studio  \cite{Prochazka2024}. 

Consistency of the preparations in the different laboratories was ensured by comparing core level spectra, LEED patterns and STM images.

\section{Results and Discussion}\label{sec2}

\subsection{Bi adsorption on Au(111)}\label{sec2.1}

Figure~\ref{fig:au111} shows reference data for clean Au(111). LEED and STM confirm the herringbone reconstruction \cite{Barth1990,Li2022}. The Au 4$f_{7/2}$ spectrum exhibits a surface core-level shift of 0.32~eV, consistent with literature values \cite{Bignardi2019,Matsui2020}. XPD patterns of the surface and bulk components reproduce the expected symmetry: nearly six-fold for the surface layer and three-fold for the bulk, as dictated by the Au(111) termination \cite{Bana2018,Bignardi2019}. The bulk component is used as reference for the crystallographic orientation.

Figure~\ref{fig:bi_phases} summarizes the structural evolution upon increasing Bi coverage. At the submonolayer regime, a $(\sqrt{37}\times\sqrt{37})$ reconstruction is observed, while for increasing Bi deposition a $(P\times\sqrt{3})$ phase appears, where $P = 5, 8, 11, 14$ describes the moiré cell size with respect to the underlying Au for increasing Bi coverages, in agreement with previous reports \cite{Jeon2009,Kawakami2017,He2019,VezzoniVicente2024}. LEED and STM confirm the high degree of order across the explored thickness range.
Bi 4$f_{7/2}$ spectra confirm cleanliness of the deposition method showing a single component at a binding energy $E_b=156.9$~eV in the multilayer regime, consistent with values found in literature \cite{Moulder:1992}. Coverage estimation, employing the XPS substrate peak attenuation, suggests for the deposited bismuth layer a thickness $< 1$~nm. Work function measurements, discussed later, give a value indicating that this is already consistent with bulk bismuth.

As reported in literature~\cite{Kawakami2017,He2019,VezzoniVicente2024}, increasing the Bi thickness leads to a progressive evolution toward a bulk-like Bi(110) termination, with the in-plane lattice aligned along the $[11\bar{2}]$ and $[1\bar{1}0]$ directions of the substrate. The XPD data confirm this trend in the multilayer regime (Fig.~\ref*{fig:bi_phases}(f)), showing qualitative agreement with simulations based on a bulk terminated Bi(110) structural model comprising six symmetry-related domains (three rotational variants and their mirror counterparts).
The resulting R-factor values are not fully satisfactory. For the pattern acquired at low kinetic energy (backscattering regime) we obtain $R=0.43$, while in the forward-scattering regime the agreement is worse, reaching $R=0.7$.
Such behavior can be rationalized in terms of the complex structure of the thick Bi overlayer described in Ref.\cite{VezzoniVicente2024} together with a sensitivity to deeper layers that increases with increasing kinetic energy. According to Refs.\cite{Lu2014,Hirayama2020}, the growth proceeds via weakly interacting bilayers stacked on a saturated interfacial Bi layer, as expected for the (110) termination. 
Away from the Au interface, the semi-layered Bi structure progressively relaxes from the $(P\times\sqrt{3})$ interfacial arrangement toward the bulk-like (110) as the film thickness increases. 
In addition to this, in the multilayer regime, the area probed by photoemission includes zones with different thicknesses, all contributing incoherently to the measured patterns.
Such a complex stacking, extensively described in the literature, differs from the simplified bulk model used in this work. Although additional buckling was introduced into the bulk-like Bi(110) structural model in an attempt to refine the agreement, this did not lead to any significant improvement. 

\subsection{Oxidation pathway of Bi films}\label{sec2.2}
\begin{figure}[h]
	\centering
	\includegraphics[width=0.4\textwidth]{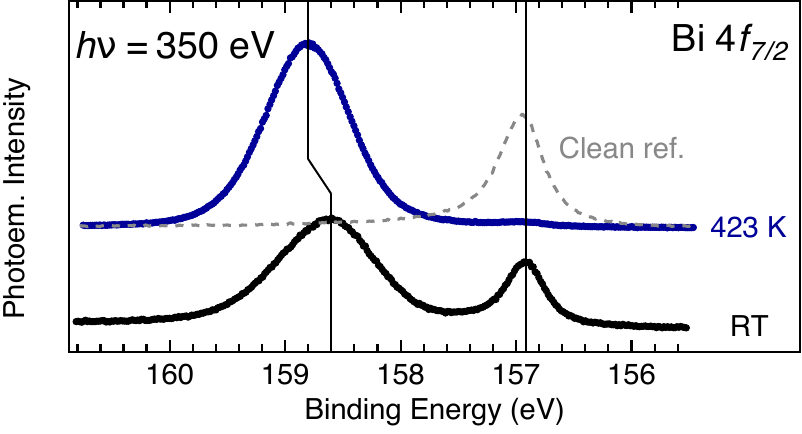}
	\caption{High resolution Bi 4$f_{7/2}$ core level spectra after exposure of Bi/Au(111) to O$_2$ at RT (black) and 423 K (blue) employing the doser, with reference (gray dashed line) for the clean Bi/Au(111). Oxidation at 423~K leads to the almost complete disappearance of the component related to metallic Bi.}
	\label{fig:xps_bi4f}
\end{figure}

In this study, two parameters were varied to simultaneously achieve an efficient oxidation while promoting structural ordering: Oxygen pressure and substrate temperature. 
Room temperature (RT) oxidation of the deposited Bi film at a background pressure of $10^{-6}$~mbar induces the gradual emergence of Bi$^{3+}$ components in the Bi 4$f$ spectrum (Fig.~S1 in \cite{SM}).
However, at such a low Oxygen pressure the oxidation is quite limited. 
A more efficient oxidation is achieved using a doser providing a local O$_2$ pressure of $10^{-3}$~mbar on the surface leading to the growth of the oxide peak at $E_b=158.6$~eV (black line in Fig.~\hyperref[fig:xps_bi4f]{\ref*{fig:xps_bi4f}}). From this point onward, all oxide preparations are carried out using this doser setup, resulting in a total exposure of $\sim10^5$~L\footnote{$1$~L$=10^{-6}$~Torr $\cdot$ s.}, unless otherwise specified, with the sample maintained at the specific temperature indicated for each case. 
Oxidation at the same pressure but at a temperature of 423~K leads to the complete disappearance of the metallic Bi and shift of the oxide peak to 158.8~eV (blue line in Fig.~\hyperref[fig:xps_bi4f]{\ref*{fig:xps_bi4f}}) indicating full conversion to Bi$^{3+}$, consistent with literature values \cite{Chen2021,Tian2021,Morgan1973}.

\begin{figure*}
	\includegraphics[width=0.8\textwidth]{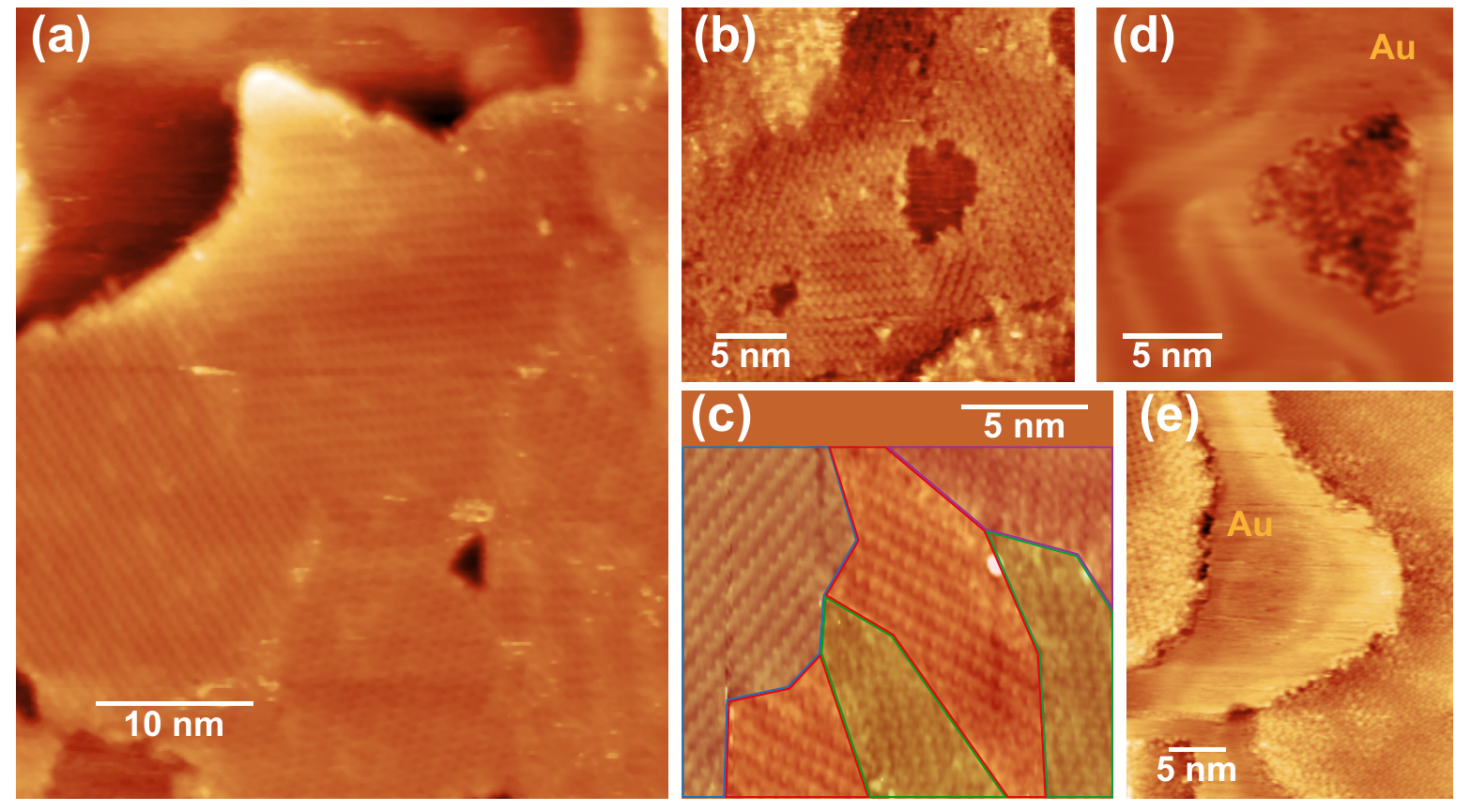}
	\caption{STM images of the thin layer of bismuth on Au(111) after oxidation at 423~K. (a) Large scale view of the surface ($I = 0.16$~nA, $V = 0.98$~V). (b) Higher resolution view of the different domains ($I = 0.24$~nA, $V = 1.00$~V) that are highlighted in (c) with different colors based on the electronic contrast produced ($I = 0.22$~nA, $V = 1.32$~V).  (d)-(e) Topography of the oxide in areas with lower coverage presenting a deformed Au herringbone reconstruction alongside oxidized bismuth. While (d) presents an island of oxide surrounded by clean gold  ($I = 0.13$~nA, $V = 0.89$~V), (e) presents an area mostly covered by the oxide ($I = 0.11$~nA, $V = 0.96$~V).}
	\label{fig:stm_150}
\end{figure*}

The total intensity of Bi 4$f_{7/2}$ photoemission (metallic + oxidized components) increases by up to a factor of 2.5 (in the case of the full oxidation at 423~K), compared to the film deposited, and the estimated thickness of the oxide layer increases to $\sim$~2.5~nm.
Such an increase in intensity  and layer thickness cannot be explained by the difference in electron mean free path inside the newly formed Bi$_2$O$_3$ layer, which would reduce the intensity, because of the lower density of Bi in the oxide, nor by photoelectron diffraction effects, since the modulations in the photoemission intensity as a function of photoemission angle are very weak. 
This behavior could be explained if an initial 3D multilayer system transforms, upon oxidation, into a quasi-2D structure, where emitters that were previously screened by the overlayers become exposed as the system evolves toward a two-dimensional configuration. 
However our STM observations, consistent with Vincente \emph{et al.}\cite{VezzoniVicente2024}, clearly show that our initial Bi surface, even in the high coverage regime, is flat and ordered as expected for metal surfaces.
Therefore we explain such an increase in intensity upon oxidation as due to a mechanism of surface segregation of Bi atoms from the bulk Au during oxidation, consistent with the known behavior of metals in Au host crystals upon oxidation, as explained by Berti \emph{et al.}\cite{Berti:2024}. 

\begin{figure*}
	\includegraphics[width=1\textwidth]{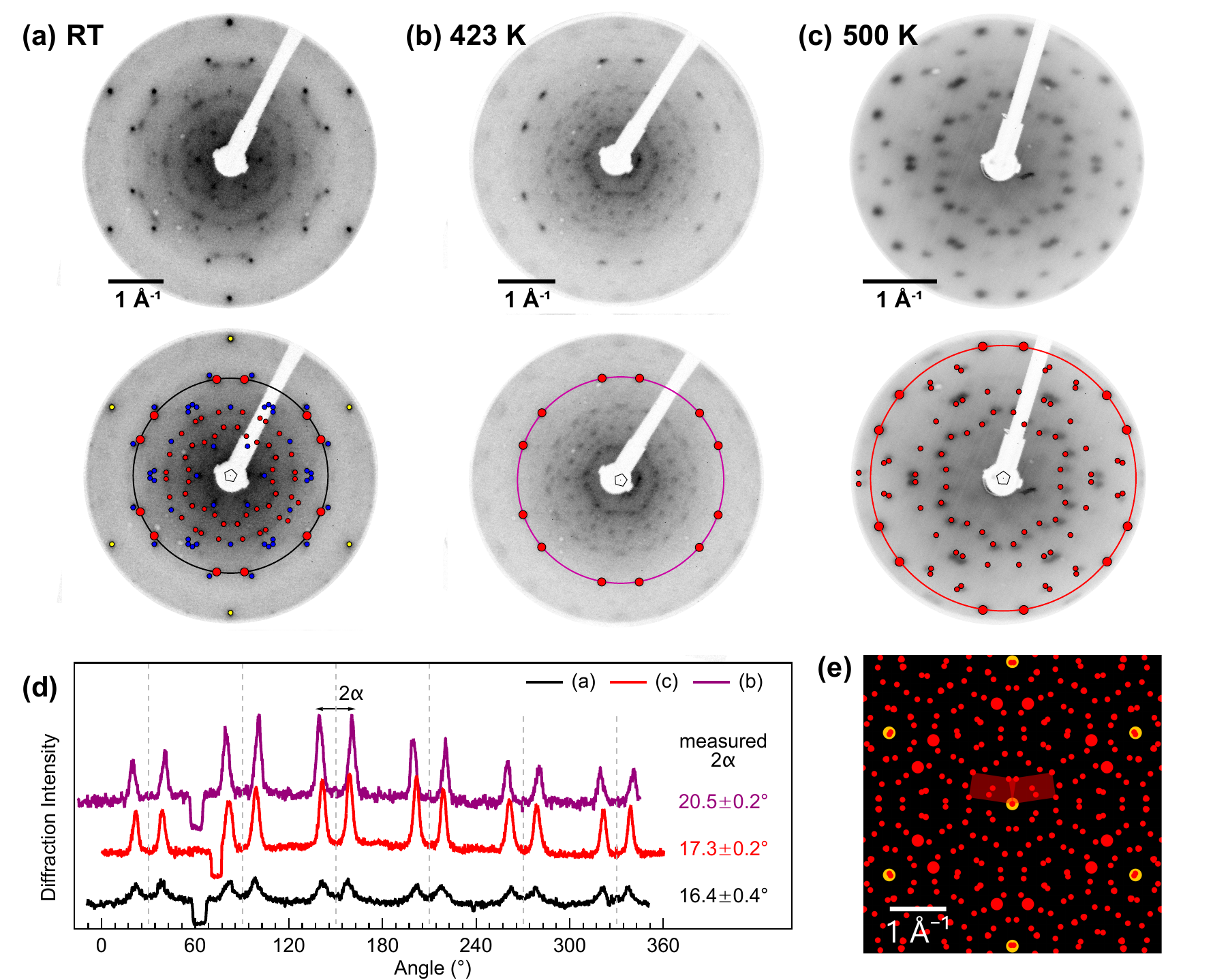}
	\caption{LEED diffraction patterns ($E_{kin} = 38$~eV) of the bismuth layer following oxidation at different temperatures. Experimental patterns (top row) are compared with corresponding simulations (bottom row). (a) Oxidation at RT, corresponding to the black spectrum in Fig. \ref{fig:xps_bi4f}. (b) Oxidation at 423~K, corresponding to blue spectrum in Fig. \ref{fig:xps_bi4f}. (c) Oxidation at 500~K, corresponding to blue spectrum in Fig. \ref{fig:annealing}(c); note that this pattern was acquired with different LEED optics, resulting in a distinct reciprocal space calibration than in (a) and (b). Diffracted spots are color-coded: yellow for the Au(111) substrate, blue for the metallic Bi$(P \times \sqrt{3})$ phase on Au(111), and red for the bismuth oxide. (d) Circular line profiles (radius $\approx 1.79$~\AA$^{-1}$) extracted from (a)-(c) along the path indicated by the respective colors. (e) Simulated diffraction pattern via kinematic theory using a matrix $\begin{pmatrix} 3.01 & 0.48 \\ 2.06 & 5.58\end{pmatrix}$, corresponding to the $\beta$-Bi$_2$O$_3$(201) termination. For clarity, two unit cells oriented at $2\alpha = 17.2^\circ$are highlighted in red.}
	\label{fig:leed_oxide}
\end{figure*}

This complex interplay between metallic bismuth, oxide formation and Au is also supported by our findings obtained by STM on the oxide formed at 423~K, shown in Fig.~{\hyperref[fig:stm_150]{\ref*{fig:stm_150}}. Large area topography (Fig.~{\hyperref[fig:stm_150]{\ref*{fig:stm_150}}(a)) shows that most of the surface is covered by a complex intermixing of differently oriented domains, as evidenced in Fig.~{\hyperref[fig:stm_150]{\ref*{fig:stm_150}}(c), while few areas still show some exposed Au surface (Fig.~{\hyperref[fig:stm_150]{\ref*{fig:stm_150}}(b)). Searching for the few portions of the surface with a clear lower coverage it is possible to observe a topography displaying a slightly distorted Au herringbone reconstruction (Fig.~{\hyperref[fig:stm_150]{\ref*{fig:stm_150}}(e)) in between islands of oxide, and the opposite in others areas (Fig.~{\hyperref[fig:stm_150]{\ref*{fig:stm_150}}(d)); however scanning parameters that highlight the herringbone reconstruction are not able to resolve the oxide topography. Similarly is also observed for oxidation at room temperature (see Fig.~S3 in \cite{SM}).

\begin{figure*}[htbp]
	\includegraphics[width=1\textwidth]{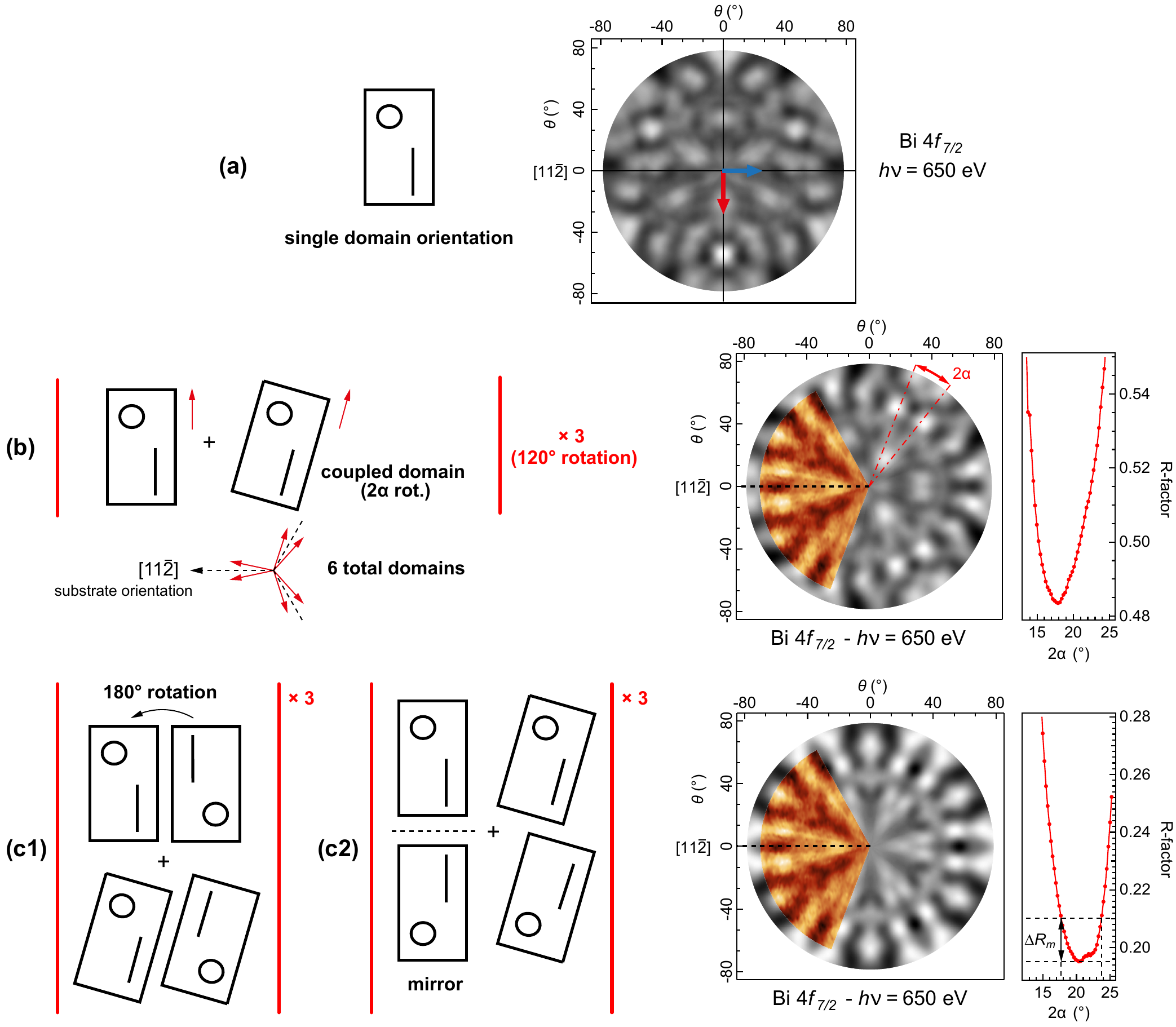}
	\caption{Left: models depicting possible combinations of $\beta$-Bi$_2$O$_3(201)$ domains on the surface, including the presence of coupled domains with a reciprocal rotation of $2\alpha$ as evidenced in LEED (Fig. \ref{fig:leed_oxide}). Right: simulated diffraction pattern associated with a model (greyscale) is compared with the experimental data (colored section) for Bi 4$f_{7/2}$ emission at $h\nu = 650$~eV ($E_{kin} = 491$~eV); the R-factor dependence on the value of the separation angle $2\alpha$ is shown on the right. (a) Model for a single orientation of $\beta$-Bi$_2$O$_3(201)$ with corresponding simulated diffraction pattern. Colored arrows represent inequivalent directions in the top view of the atomic model presented in Fig.~\ref{fig:201_cell}. While model (b), consisting of 6 unique domains, gives $R_m=0.48$, model (c1) (or equivalently (c2)), with 12 unique domains, produces a significantly improved agreement with the experimental data with $R_m=0.20$. The confidence interval for the value $R_m$ is reported.}
	\label{fig:xpd}
\end{figure*}

LEED patterns acquired after oxidation at different temperatures are shown in Fig.~\ref{fig:leed_oxide}. These display numerous faint and broad diffraction spots, consistent with the small oxide domains observed by STM. Despite this complexity, twelve bright spots at $k_{\parallel} \approx 1.78$~\AA$^{-1}$ are systematically observed, with only minor variations in their angular separation $2\alpha$ (Fig.~\ref{fig:leed_oxide}(g)) as a function of oxidation temperature.
The persistence of these features suggests only a partial long-range order, consistently with what is observed in STM. 
Radial position and symmetry of the LEED spots are compatible with the $(201)$ surface termination of the $\beta$-Bi$_2$O$_3$ phase (Fig.~\ref{fig:201_cell}), already observed in previous reports\cite{Tian2021,Saito2022,Jiang2020}. 
Fig.~\ref{fig:leed_oxide}(a)-(c) shows the acquired diffraction patterns for three relevant oxidation temperatures,  and the results obtained by kinematic simulations for the  Au(111) (yellow circles), $(P \times \sqrt{3})$ phase for metallic Bi  (blue circles), and the $\beta$-Bi$_2$O$_3 (201)$ terminations (red circles), respectively.
In particular, the twelve diffraction spots at $\approx 1.79$~\AA$^{-1}$, marked by large red circles, can be rationalized by rotational and/or mirror domains of the $(201)$ surface of the oxide, comprising unit cells rotated by $\pm\alpha$ relative to the $[11\bar{2}]$ directions of the three-fold Au(111) substrate. A simulation of the LEED pattern, including the diffraction spots not visible in experimental data, is shown for clarity in Fig.~\ref{fig:leed_oxide}(e), where two differently oriented unit cells are highlighted in red around the $(0,0)$ spot.

The coexistence of patterns stemming from both metallic Bi and the oxide layer after oxidation at RT (Fig.~\ref{fig:leed_oxide}(a)) is consistent with the partial oxidation observed via XPS. This is further supported by their complete disappearance upon oxidation at 423~K, in agreement with XPS data (blue spectrum in Fig.~\ref{fig:xps_bi4f}). The pattern becomes clearer, even in the lower-order reflections, for oxidation at 500~K; interestingly, this pattern is identical to the one obtained after oxidation at 1~mbar at RT (see Fig.~S2 in \cite{SM}).
The coexistence of multiple ordered domains with various azimuthal orientations, as evidenced by LEED and STM, results in a lack of pronounced XPD features for low kinetic energy photoelectrons ($E_{kin} \approx 100$~eV) in the backscattering regime. Conversely, in the more bulk-sensitive forward-scattering regime — accessed via higher photon energies ($E_{kin} = 491$~eV) — the acquired pattern exhibits clearly defined features, as demonstrated in Fig.~\ref{fig:xpd} for the oxide prepared at 423~K.  

This diffraction pattern further aligns with the $\beta$-Bi$_2$O$_3$(201) termination as a recurring local structural motif, consistently with LEED observations.
Fig. \ref{fig:201_cell} shows the structural model used for the simulations, that include emitters from the first three Bi layers, while deeper layers do not significantly affect the pattern (R-factor variation $\lesssim 0.01$ upon inclusion).

\begin{figure}[htbp]
	\centering
	\includegraphics[width=0.35\textwidth]{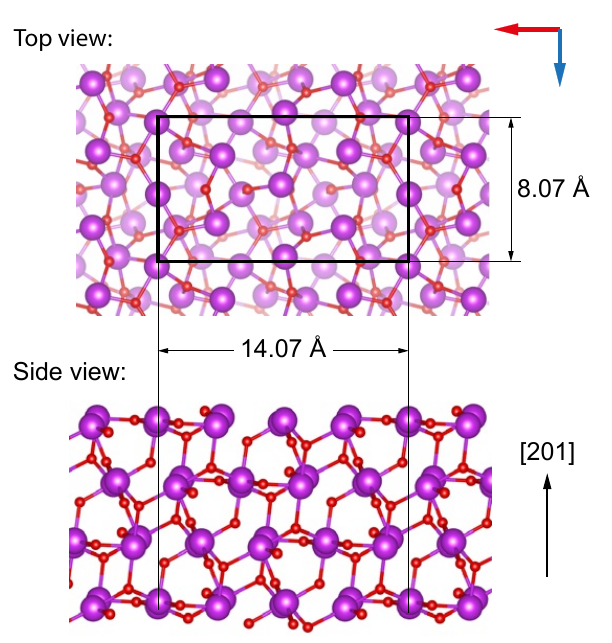}
	\caption{Atomic model of the $\beta$-Bi$_2$O$_3(201)$ termination employed for XPD simulation; Purple - Bi, Red - O.  Colored arrows represent inequivalent directions in the top view of the atomic model here as well as in the diffraction pattern shown in Fig.~\ref{fig:xpd}(a).}
	\label{fig:201_cell}
\end{figure}

While simulations assuming a single domain (Fig.~\ref{fig:xpd}(a)) fail to reproduce the experimental pattern, the inclusion of six domains (Fig.~\ref{fig:xpd}(b)) — specifically two domains rotated by $\pm \alpha$ with respect to the $[11\bar{2}]$ direction (as per Fig.~\ref{fig:leed_oxide}) and their threefold symmetric equivalents — improves the agreement with the measured data. However, a quantitative comparison still yields an unsatisfactory $R$-factor of 0.48.  
Inclusion of additional mirrored domains (equivalently, 180$^\circ$-rotated), as shown in (Fig.~\ref{fig:xpd}~(c1)-(c-2)), substantially improves the R-factor and enables an accurate reproduction of the experimental pattern, yielding $R = 0.20$.

Following an approach analogous to the Pendry analysis of LEED I–V curves \cite{Pendry1980}, the uncertainty in the minimum R-factor, $R_m$, can be estimated as $\Delta R_m = \sqrt{2/N_p} R_m$, where $N_p$ is the number of well-separated diffraction peaks. Estimating $N_p$ from the azimuthal XPD scans at different polar emission angles \cite{Bana2018,Bignardi2019} gives $\Delta R_m = 0.015$. This constrains the angular separation to $2\alpha = 20.7 \pm 3.0^\circ$, in agreement with the LEED analysis shown before.

\subsection{High-temperature restructuring}\label{sec2.3}

In order to gain insight into the ordering dynamics of the oxide and its stability upon annealing, we performed temperature-dependent XPS measurements in UHV on samples oxidized at room temperature presented in Fig.~\ref{fig:annealing}.  

\begin{figure*}[htbp]
	\centering
	\includegraphics[width=\textwidth]{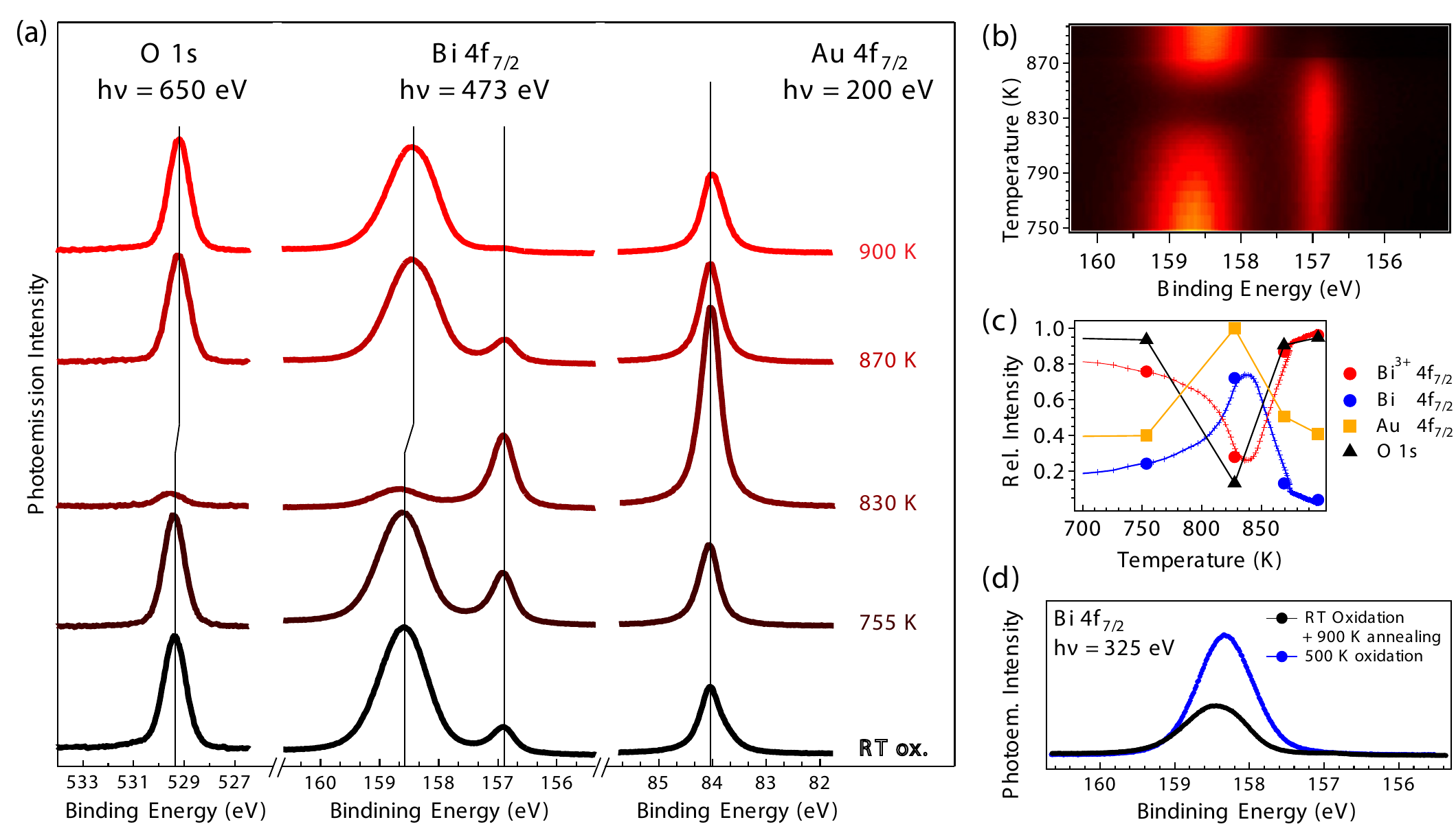}
	\caption{XPS evidence of the complex restructuring sequence following annealing of the bismuth oxide layer prepared with RT oxidation. (a) Temperature resolved spectra of the Bi 4$f_{7/2}$ core level acquired at $h\nu = 325$~eV, displaying significant peak modulation due to surface reordering. The obtained figure is the relevant part of a larger scan that starts from RT and reaches 900~K. (b) High resolution spectra for O 1$s$ ($h\nu = 650$~eV), Bi 4$f_{7/2}$ ($h\nu = 473$~eV) and Au 4$f_{7/2}$ ($h\nu = 200$~eV) core levels at different steps of the annealing ramp. Each spectrum was acquired after the sample had cooled down to RT. (c) Relative photoemission intensity as a function of temperature for the relevant spectral lines. (d) High resolution Bi 4$f_{7/2}$ core level spectra comparison after the annealing ramp and after 500~K oxidation of the deposited Bi layer.}
	\label{fig:annealing}
\end{figure*}

For this preparation (black spectra in Fig.~\ref{fig:annealing}(a)), the system presents a small amount of metallic bismuth ($E_b=156.87$~eV), and the Au 4$f_{7/2}$ core level retains the low-binding-energy component ($E_b=83.72$~eV}) characteristic of a clean surface contribution (see  Fig.~S4 in \cite{SM} for fitting details). 
This suggests that portions of the Au substrate remain uncovered by the oxide prepared under these conditions, consistent with observations obtained by STM (Fig.~S3 in \cite{SM}) and ARPES (see the following section, Fig.~\ref{fig:ARPES_BiOx}(a)).

Upon annealing to 755~K, a loss of oxidized Bi spectral weight is observed, accompanied by the disappearance of the low binding energy component of the Au 4f$_{7/2}$.
Fig.~\ref{fig:annealing}(b) shows in colour scale the evolution as a function of temperature of the photoemission intensity in the Bi 4$f_{7/2}$ region. 
Fig.~\ref{fig:annealing}(c)  displays the relative intensity of the relevant spectral lines as a function of temperature, defined as the ratio of the normalized area under the fitted component to its measured maximum value.
At 830~K, the oxide intensity drops drastically as metallic Bi reappears, accompanied by a nearly threefold increase in the Au spectral line intensity. The latter reaches a value comparable to that of the surface component in clean Au(111) (see Fig.~S4 in \cite{SM}). 
This behavior suggests the loss of the oxide layer and the reestablishing of a clean Bi$\slash$Au system. However, the absence of the low BE component of the Au 4$f_{7/2}$ foretells that the situation is more complex than a mere \emph{fall apart} of an oxide upon annealing.
Surprisingly, in fact, when reaching temperatures above 870~K and without supplying any O$_2$, the reappearance of the oxide is observed.
The Bi$^{3+}$ component becomes the predominant one and appears at a slightly lower binding energy than the pristine oxide produced at RT ($E_b=158.4$~eV), concurrently, the metallic Bi component nearly disappears, while the Au signal undergoes an abrupt loss. This provides clear evidence that the Au substrate becomes buried beneath the oxide layer.

Oxidation performed directly at 500~K, while giving the same ordered structure of the oxide as seen by LEED in Fig.~\ref{fig:leed_oxide} after the annealing ramp, results in an even lower binding energy value ($100$~meV as seen in Fig.~\ref{fig:annealing}(d)). Moreover, subsequent annealing does not produce the evolution sequence displayed above for the RT oxidation. These combined observations suggest that annealing of the RT-prepared oxide drives the system toward a very similar chemical state achieved by direct high-temperature (500~K) oxidation.

Such a complex evolution can be rationalized as follows.
For $T \lesssim$~830~K, oxygen redistributes from the near-surface region throughout the Bi film. The concomitant decrease in Bi 4$f$ oxide-component  and O 1$s$ intensity, together with the increase in metallic Bi and Au signals, is consistent with Au (and metallic Bi) surface segregation, likely filling vacancies generated by oxygen redistribution upon annealing and then eventually migrating to the surface. 
At higher temperature, the system undergoes a structural transformation toward a more stable Bi$_2$O$_3$ polymorph. In this regime, oxide formation becomes energetically favored over Au–Bi alloying at the surface, leading to a renewed increase in oxidized Bi and O 1$s$ intensity and a corresponding suppression of metallic Bi and Au contributions.
All this is consistent with what is observed in ARPES and presented in the following section.


\subsection{Electronic structure}\label{sec2.4}

In order to better understand the modifications on the electronic structure, we performed ARPES measurements that are presented in Fig. \ref{fig:ARPES_Au} and \ref{fig:ARPES_BiOx}.
Bi deposition in the $\left(\sqrt{37}\times\sqrt{37}\right)R25.3^\circ$ regime, leads to the disappearance of the Au surface state (SS) and the new larger periodicity on the surface leads to a folding of the band structure that becomes clearer with longer Bi deposition leading to the multilayer regime. Such umklapp scattering of the Au $sp$ band indicated by the black arrows in Fig. \ref{fig:ARPES_Au}(c) has been observed and explained in the similar Sb$\slash$Au(111) system in Ref.\cite{Hu:2025}. 

\begin{figure*}[htbp]
	\centering
	\includegraphics[width=1\textwidth]{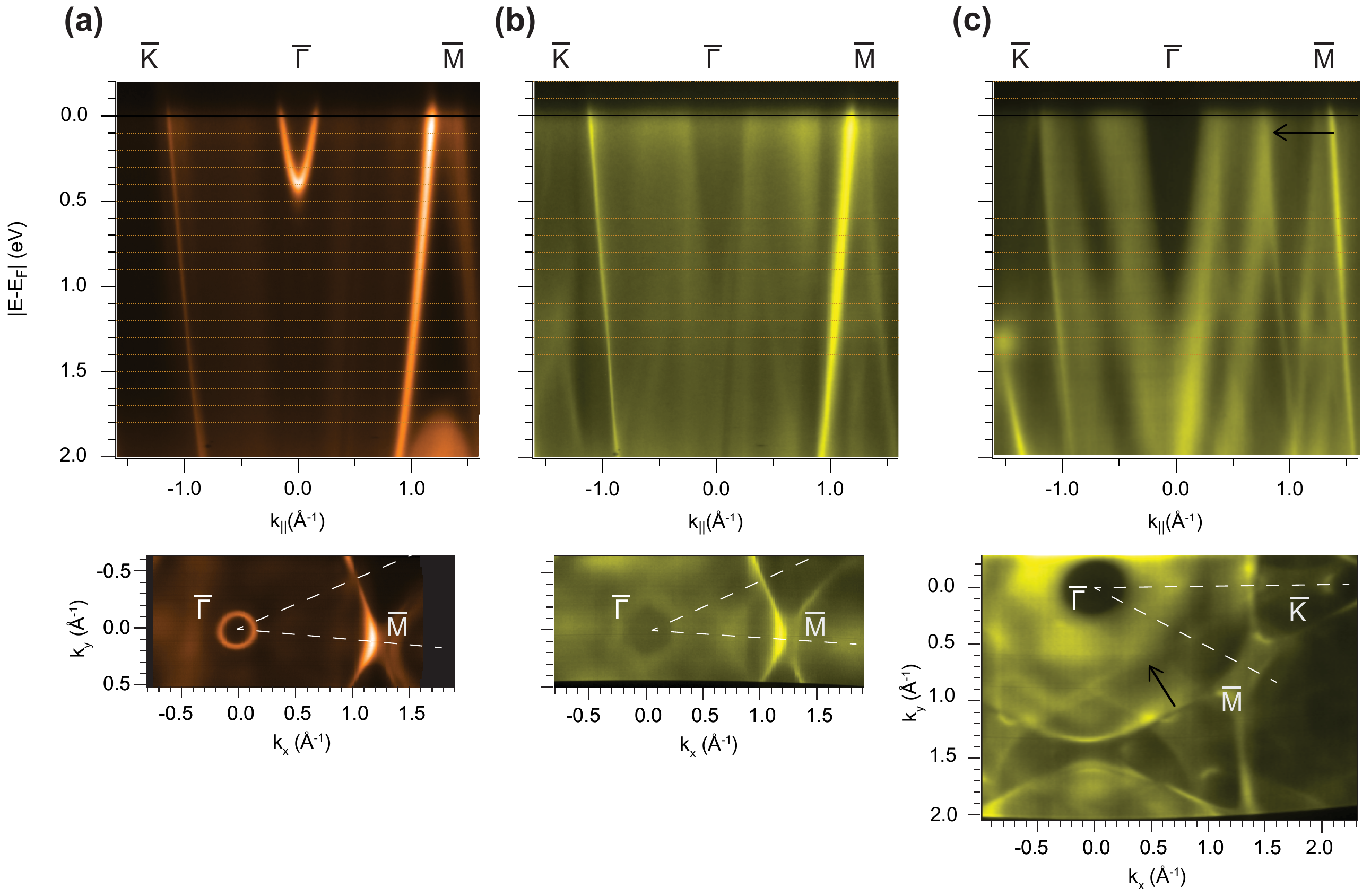}
	\caption{Energy dispersion as a function of momentum along the ${\overline K}{\overline \Gamma}{\overline M}$ direction (top) as depicted in the measured Fermi surfaces (bottom). ARPES measurements in (a) are acquired on clean Au(111) while (b) and (c) on Bi$\backslash$Au(111) in the submonolayer and multilayer regimes, respectively. Spectra in (a) and (b) have been acquired at h$\nu=32$~eV while the ones in (c) at h$\nu=70$~eV in order to probe a larger area in k-space, black arrow indicates folding of the band structure due to umklapp scattering.}
	\label{fig:ARPES_Au}
\end{figure*}

Consistent with the process of Bi segregation suggested by XPS (Fig. \ref{fig:xps_bi4f}) and by STM (Fig. \ref{fig:stm_150}(d)-(e)}), oxidation at 423~K of the submonolayer regimes recovers a surface state that resembles the one from Au(111), as shown in Fig. \ref{fig:ARPES_BiOx}. However, the higher disorder of the surface with respect to the pristine one causes a broadening that hinders resolving its characteristic Rashba splitting. A similar result is obtained for oxidation at 500~K (data not shown).
ARPES is not able to determine if the observed SS actually lies at the interface of Au and Bi$_2$O$_3$ or arises from the areas of gold that are left uncovered by the oxidation process.
Nevertheless, due to its origin as Shockley state, and therefore its characteristic fragility in presence of contamination on the surface, it is reasonable to think that the latter is the case.
This is also consistent with the distorted herringbone reconstruction observed in Fig.~{\hyperref[fig:stm_150]{\ref*{fig:stm_150}(d)-(e)}}. 

\begin{figure*}[htbp]
	\centering
	\includegraphics[width=1\textwidth]{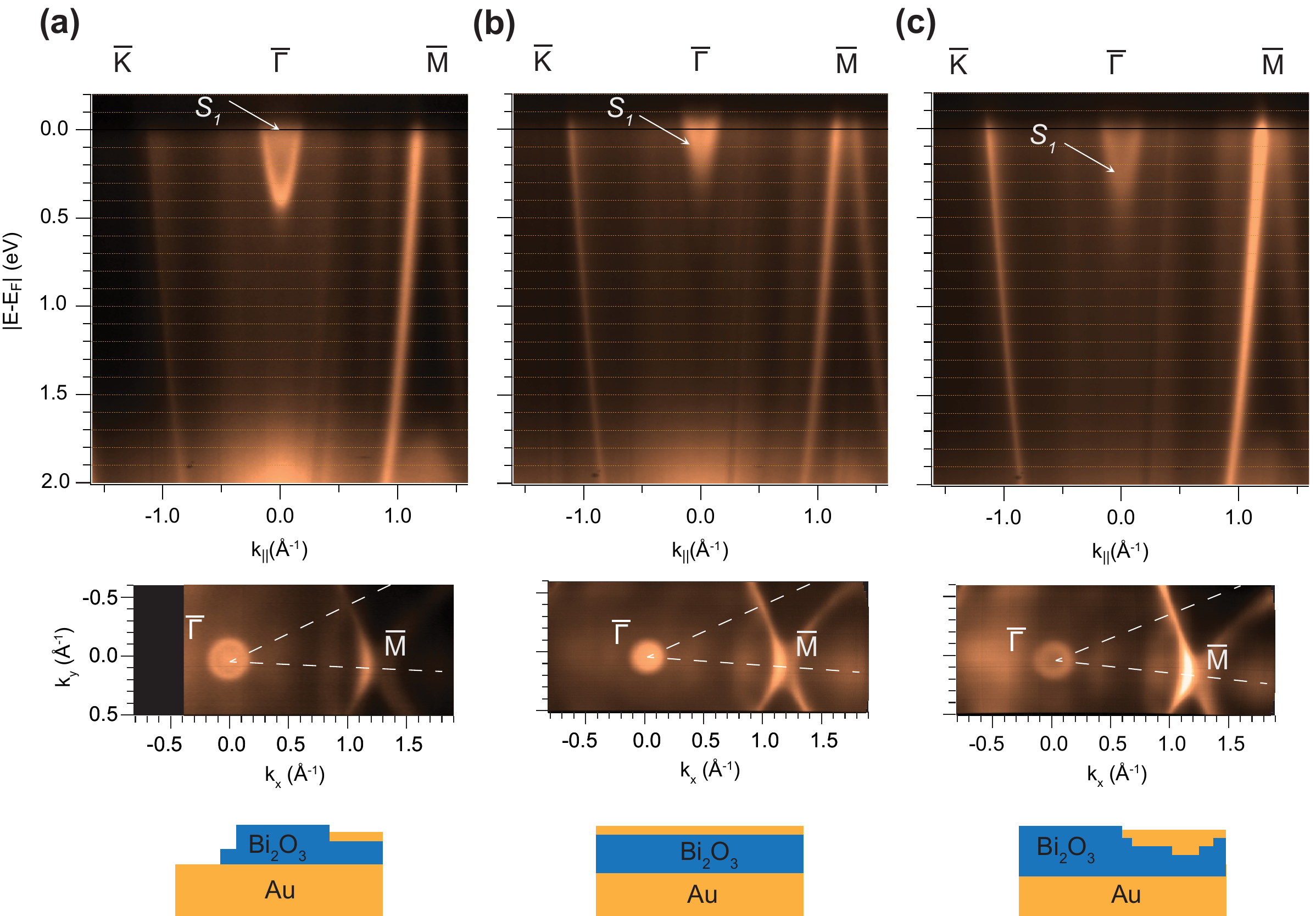}
	\caption{Energy dispersion as a function of momentum along the ${\overline K}{\overline \Gamma}{\overline M}$ direction (top) as depicted in the measured Fermi surfaces (middle). ARPES measurements on Bi$_2$O$_3$$\backslash$Au(111) prepared at 423~K (a) and after annealing at 820~K (b) and 890~K (c), respectively,  acquired at h$\nu=32$~eV. White arrow indicates the extra electron pocket $S_1$ appearing and shifting upon annealing at the expenses of SS. The proposed models for bismuth oxide and Au distribution are presented at the bottom.}
	\label{fig:ARPES_BiOx}
\end{figure*}

What is more surprising is the appearance of an extra electron pocket ($S_1$) at $\overline\Gamma$, \emph{inside} the SS (Fig.~\ref{fig:ARPES_BiOx}) that becomes more visible upon annealing at the expenses of the gold SS.
S1 cannot be reconciled with the specific (201) termination of the oxide, though a minimum in its conduction band at $\Gamma$ is predicted by calculations published in literature \cite{Xiong2025}, that show a thickness dependent indirect band-gap ranging from 2.2~eV for thin layers ($< 1$~nm) to 1.6~eV for a thickness of about 2~nm, with the difference attributed to quantum confinement effects.

The origin of this state can be traced back to the complex evolution indicated by the observed high-temperature restructuring in XPS (Fig.~\ref{fig:annealing}), which results in a redistribution of Au and oxygen inside the layer, and eventually the segregation of Au on the surface,  resulting in the formation of Au ponds.
The formation of Shockley states in quantum well structures of thin metals has been extensively studied theoretically, and simple models for their formation have been proposed\cite{Milun2002}, even if observation of these are scarce in recent literature. It was shown by Varykhalov \emph{et al.}\cite{Varykhalov2020} that the SS originating from thin Au films on W(110) experiences a shift to higher binding energy of around 100~meV as the film thickness increases, attributed to strain caused by the substrate, while the case of Au/Ag(111)\cite{Palomares2002} leads to a comparable result, with a gradual shift of the observed peak intensity towards the pristine Au(111) value. A very similar behavior is observed here for $S_1$, supporting the attribution of the electron pocket to Au on the surface. 

The evolution upon annealing observed by XPS (Fig.~\ref{fig:annealing}) and ARPES can be reconciled thanks to the gathered insights (see bottom line of Fig.~\ref{fig:ARPES_BiOx}). 
After intermediate temperature oxidation (433~K) the oxide layer does not uniformly cover the surface, as pointed out previously. This implies that the pristine Au(111) SS is visible, along with the pocket $S_1$ that can be attributed to thin Au \emph{puddles} on the oxide. 
After annealing at 830 K, XPS reveals a significant decrease in the oxide signal from the surface layers, accompanied by a striking increase in the Au 4$f_{7/2}$  signal. This behavior is consistent with the formation of a largely uniform, thin Au layer on top of the bismuth oxide, which leads to the disappearance of the surface state (SS) of pristine Au(111) and the appearance of the S1 pocket with increased intensity.

Upon further annealing up to 900~K the Bi 4$f_{7/2}$ oxide component re-gains most of the lost intensity at the expenses of Au 4$f_{7/2}$, compatibly with the coexistence on the surface of bismuth oxide and thicker Au areas on top of it, as suggested by the shift of $S_1$ towards higher binding energy.

Work function measurements presented in Table \ref{tab:workfunction} show a substantial decrease upon oxidation, dropping from 4.1 eV for Bi/Au(111) to $\sim 3.4$~eV, depending on the specific preparation of the Bi$_2$O$_3$/Au(111) system. The oxide film prepared at 423 K exhibits the lowest work function, which correlates with the highest binding energy (BE) observed for the Bi $4f_{7/2}$ core level (see Fig. S5 in \cite{SM}). Conversely, the system prepared at 503 K yields the highest measured $\Phi$, concurrently with the lowest Bi $4f_{7/2}$ BE. Upon further annealing, the work function decreases again, a trend that tracks the shift in the Bi $4f_{7/2}$ (and O $1s$) BEs . Such behaviour reflects the complex segregation dynamics occurring at different annealing temperatures, as previously described. For contact engineering applications, the oxide prepared at 423 K represents the most relevant case; here, the presence of metallic species within the oxide lattice appears to be minimized, coinciding with a minimized $\Phi$ and a maximized Bi $4f_{7/2}$ binding energy. Other temperatures lead to different concentration of interlayer metallic species (such as Au or oxygen vacancies), and therefore different screening effects that on average reduce the oxidation state, ultimately leading to a higher $\Phi$.



\begin{table}[!h]
	\centering
	\begin{tabular}{|ccc|}
		\hline
		\textbf{Surface    } & \textbf{Preparation} &    \hspace{0.5cm}$\Phi$~(eV)\hspace{0.5cm} \\
		\hline
		Au(111) & clean & 5.4 \\
		Bi/Au(111) & after evaporation  &4.1 \\
		Bi$_2$O$_3$/Au(111) &  503~K oxidation & 3.8\\
		&+ annealing 673~K & 3.6\\
		&+ annealing 923~K & 3.5\\
		Bi$_2$O$_3$/Au(111) &  423~K oxidation & 3.4\\
		\hline
	\end{tabular}
	\caption{Experimentally measured work functions. All the oxides have been prepared using the oxygen doser as explained in the Methods section.}

	\label{tab:workfunction}
\end{table}


\section{Conclusions and Outlook}\label{sec3}

We have provided a comprehensive, multi-technique investigation detailing the growth, oxidation pathway, and thermal stability of bismuth oxide thin films on the Au$(111)$ surface. By synergistically combining synchrotron-based spectroscopy (XPS, XPD) with real-space microscopy (STM) and LEED, we have successfully identified the structure and unraveled the electronic evolution of this complex interface.
Diffraction measurements demonstrate that controlled oxygen exposure leads to the formation of an ordered, heterogeneous oxide layer whose local arrangement is consistent with the $(201)$ surface termination of the metastable $\beta-Bi_{2}O_{3}$ polymorph. This phase accommodates the lattice mismatch with the substrate by organizing into twelve rotationally and mirror-symmetric domains, providing a robust structural benchmark for ultra-thin bismuth oxide films. 

Crucially, temperature-dependent measurements revealed a non-trivial restructuring mechanism driven by thermal annealing. Rather than a simple thermal decomposition, the system undergoes a dynamical interplay governed by oxygen redistribution and the competitive segregation of Au and metallic Bi atoms. At elevated temperatures (above 870 K), this complex migration leads to a reconfiguration where the $\text{Au}$ substrate becomes buried beneath a stable oxide overlayer, accompanied by the formation of distinct surface electronic states, as captured by ARPES. 

Beyond filling a fundamental gap regarding the structural landscape and wetting properties of the ultra-thin Bi-oxide/Au(111) system, these insights offer significant potential for advanced technological applications. Importantly, the growth protocol and the thermal reconfiguration processes described herein rely on direct deposition and controlled oxidation pathways that are intrinsically scalable to large-scale processing, making this platform highly compatible with existing semiconductor manufacturing technologies. 
Furthermore, the ability to precisely tune the work function and manipulate the surface electronic properties through targeted oxidation and thermal annealing provides a powerful degree of freedom for interface engineering. The dramatic reduction of the work function upon oxidation — reaching values as low as $\sim 3.4\text{ eV}$ — establishes this system as a highly promising candidate for contact engineering in next-generation nanoelectronics. Specifically, utilizing ultra-thin $\beta-\text{Bi}_2\text{O}_3$ layers as an interlayer on gold electrodes could drastically lower the Schottky barrier heights when interfacing with two-dimensional ($\text{2D}$) semiconductors, such as transition-metal dichalcogenides ($\text{TMDs}$). By acting as an effective screening barrier against metal-induced gap states while potentially enabling efficient charge-transfer doping, this platform points toward a technologically viable pathway for achieving ultra-low contact resistance in $\text{2D}$ material-based field-effect transistors and optoelectronic devices. Future efforts will be directed toward the direct integration of these optimized, scalable oxide templates with $\text{TMD}$ monolayers to validate their performance in operational device architectures.



\section*{Acknowledgments}
We acknowledge support from the staff of Elettra-Sincrotrone Trieste, in particular of Dr. Paolo Lacovig. 
Al.Ba. acknowledges the financial support from the Italian National Quantum Science and Technology Institute (PNRR MUR project-PE0000023-NQSTI).
A.T. acknowledges support from Collegio Universitario "Luciano Fonda". P.H acknowledges financial support by the Independent Research Fund Denmark  (Grants No. 1026-00089B and 4258-00002B) and by the Novo Nordisk Foundation (Grant number NFF23OC0085585).

\section*{Data availability}
The data that support the findings of this study are available from the corresponding authors upon reasonable request.

\FloatBarrier

\section{Supplementary Notes}
\subsection*{Oxidation at room temperature and low pressure}
Oxidation in ultra high vacuum range, specifically at pressure of $1.5\times10^{-6}$~mbar,  leads to the emergence of new components in the Bi 4$f$ binding energy region compatible with Bi$^{+3}$ species. However the process is not as efficient as when using the oxygen doser described in the main text. This is presented in Fig.~\ref{fig:xps_bi4f_LP}, where spectra acquired while dosing oxygen in background pressure at room temperature are shown.
 
\begin{figure}[htbp]
	\includegraphics[width=0.8\columnwidth]{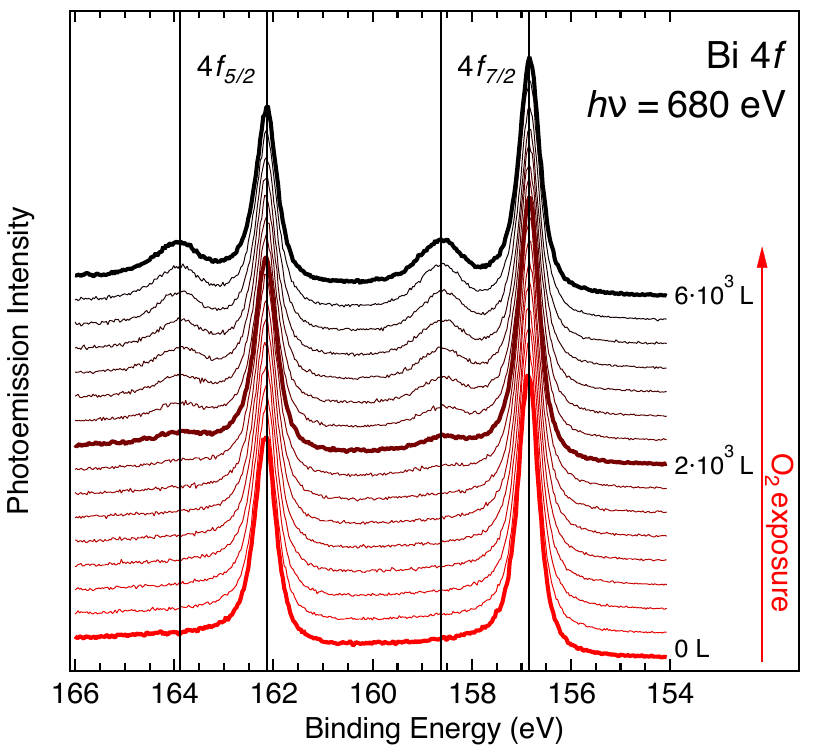}
	\caption{Bi 4$f$ core level spectra of Bi/Au(111) as a function of O$_2$ exposure at RT at low pressure. New components appearing at higher binding energy that are assigned to oxidized Bi$^{+3}$ species.}
	\label{fig:xps_bi4f_LP}
\end{figure}

\subsection*{Oxidation at room temperature and high pressure}

Oxidation at room temperature in a dedicated high-pressure cell (1 mbar O$_2$) yields a diffraction pattern—shown in Fig.~\ref{fig:leed_1mbar}(a)—analogous to the one obtained at higher temperature (500~K) and lower pressure using the oxygen doser. This underscores the interplay between temperature and pressure in the oxidation process. Such equivalence is further supported by the line profile analysis in Fig.~\ref{fig:leed_1mbar}(b), which indicates that the domain rotation angle ($\alpha$) remains consistent between the two preparation methods within experimental uncertainty. 

\begin{figure}[htbp]
	\includegraphics[width=\columnwidth]{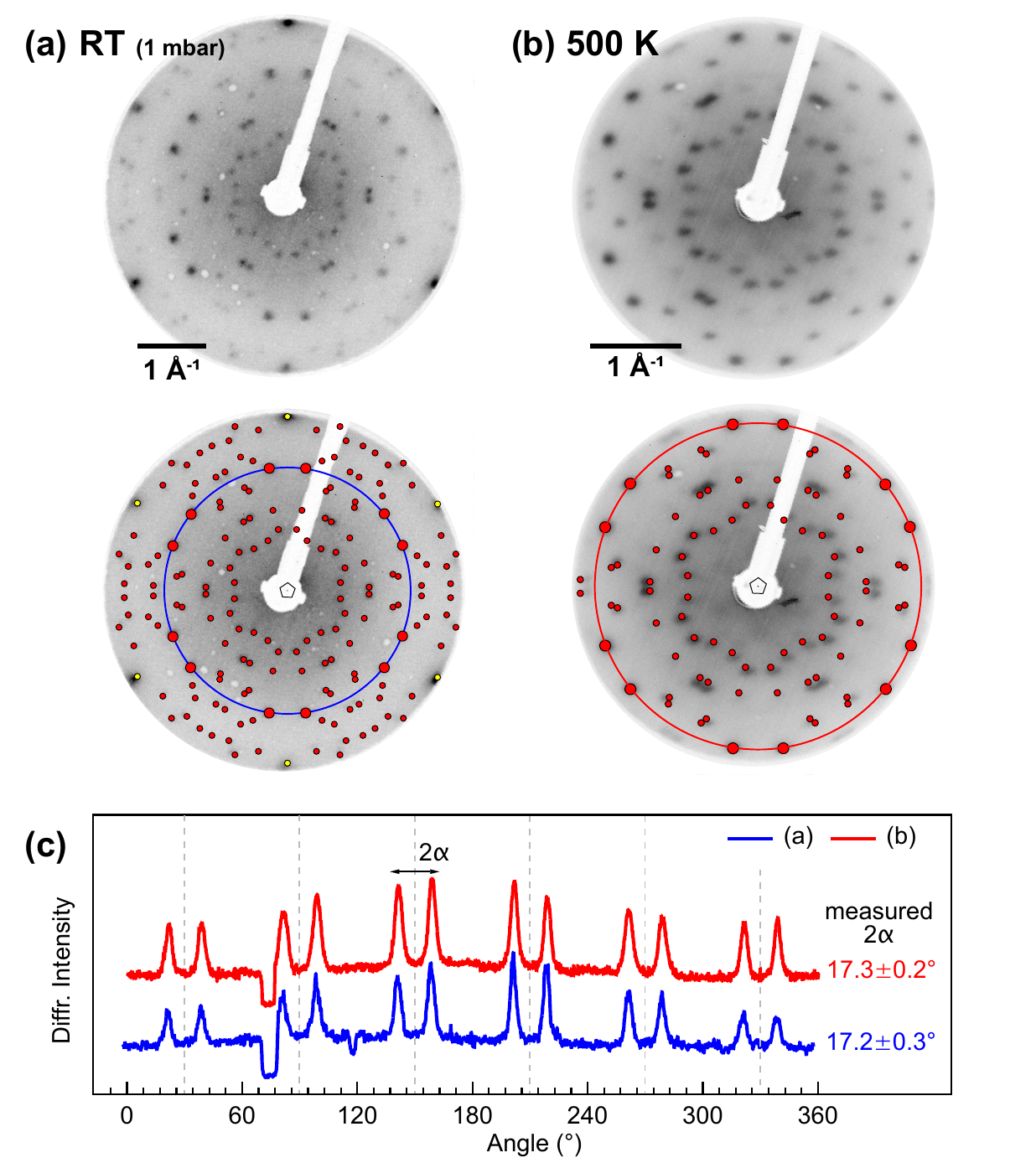}
	\caption{LEED patterns ($E_{kin} = 38$~eV) of Bi/Au(111) following oxidation at different temperatures and pressures. Experimental data (top) are compared with simulations (bottom). (a) RT oxidation at 1~mbar in the high pressure cell.  (b) Oxidation at 500~K using the doser; note the different reciprocal space calibration due to distinct LEED optics. Diffracted spots are color-coded: yellow for the Au(111) substrate, and red for the bismuth oxide. Simulation parameters are identical to those used in Fig.~5 of the main text. (c) Circular line profiles (radius $\approx 1.78$~\AA$^{-1}$) extracted from (a) and (b) along the path indicated by the respective colors.}
	\label{fig:leed_1mbar}
\end{figure}

Similarly, surface topography shown in Fig.~\ref{fig:STM_RT} reveals analogous features for oxidation at both RT, and 423~K (main text Fig.~4). These findings, in conjunction with the data presented in the main text, confirm the stability and structural nature of the oxide formed on Au(111). They suggest that the differences between the distinct preparations arise from the density of oxygen vacancies—as evidenced by high-resolution XPS— and degree of long range order rather than from fundamental structural modifications.

\begin{figure}[htbp]
	\includegraphics[width=\columnwidth]{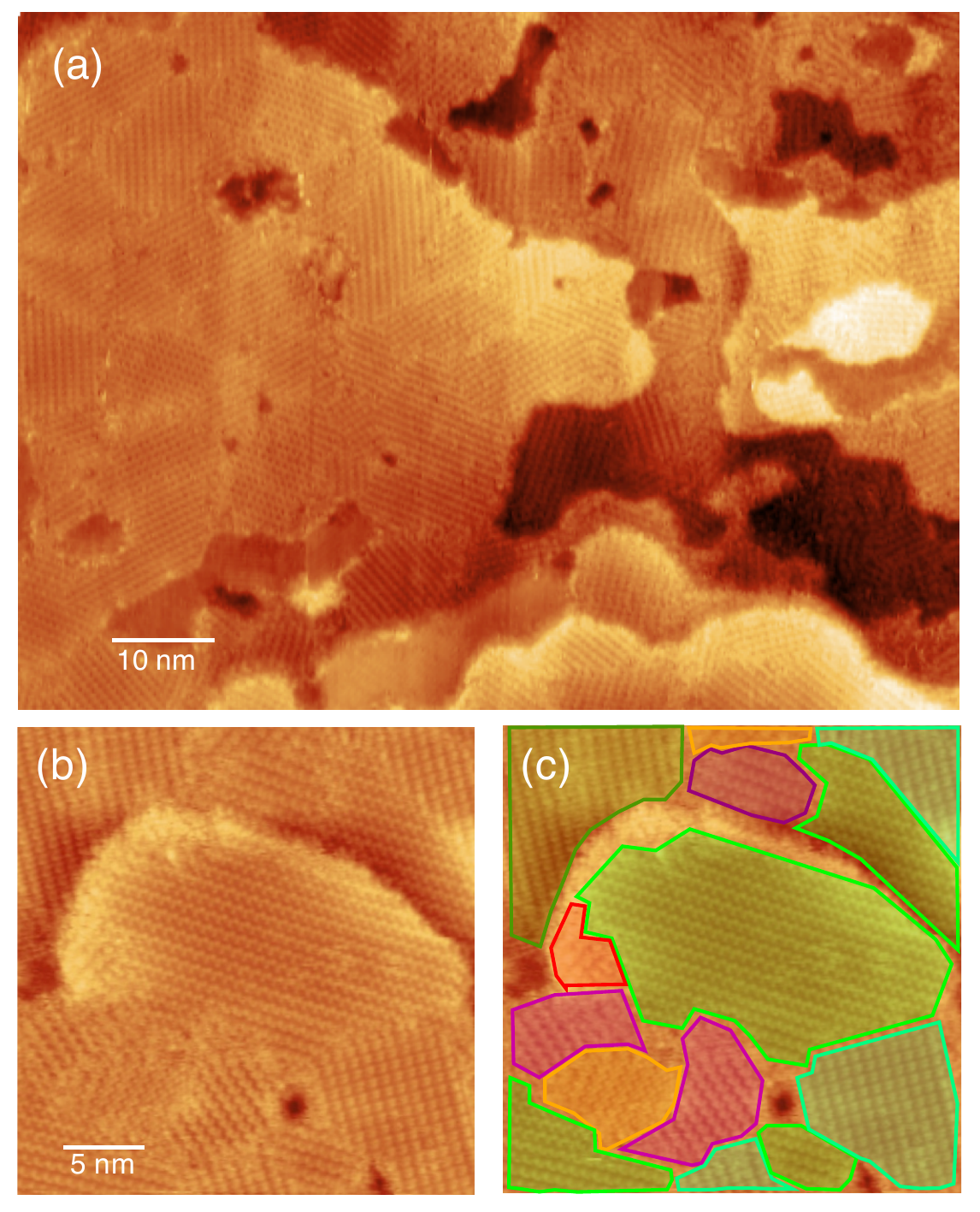}
	\caption{STM images of the thin layer of bismuth on Au(111) after oxidation at RT using the doser. (a) Large scale view of
the surface (I = 0.16 nA, V = 0.98 V). (b)-(c) Identification of distinct domains, color-coded based on their
electronic contrast (I = 0.22 nA, V = 1.32 V).}
	\label{fig:STM_RT}
\end{figure}

\subsection*{Temperature dependent XPS spectra}

Detailed insights into the Au $4f_{7/2}$ spectra, acquired at a photon energy of 200 eV upon annealing, are provided in this section. Figure~\ref{fig:Au4f_TD} shows the spectral decomposition based on Doniach-Šunjić line shapes, where the bulk and surface/2D components are represented in red and yellow, respectively. 

\begin{figure*}[htbp]
\includegraphics[width=0.7\textwidth]{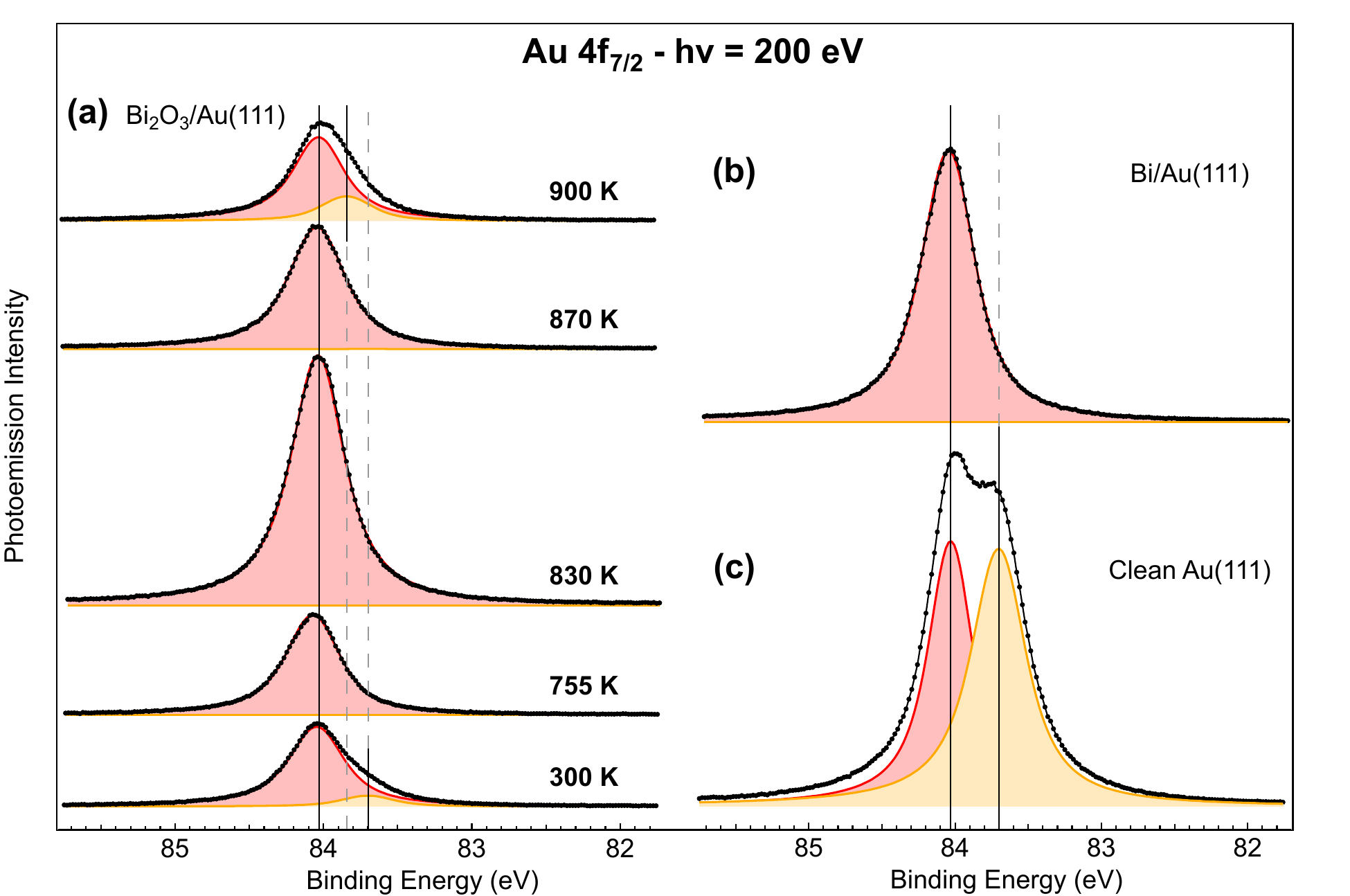}
	\caption{Au 4$f_{7/2}$ core level spectra acquired at h$\nu=200$~eV of $\beta$-Bi$_2$O$_3$/Au(111) after annealing at different temperatures. In Red is marked the bulk related component while in Yellow the surface/2D related one. On the right side is shown the clean Au(111) surface reference.}
	\label{fig:Au4f_TD}
\end{figure*}

 To support the interpretation of the complex segregation dynamics and their impact on the measured work function and core levels, Fig.~\ref{fig:CL_vs_wf} displays high-resolution spectra acquired for the Au $4f_{7/2}$, Bi $4f_{7/2}$, and O $1s$ core levels for the oxide prepared at different temperatures, alongside the corresponding work function ($\Phi$) values. As discussed in the main text, the lowest $\Phi$ is observed for the oxide prepared at 423 K. This state is characterized by the highest Bi $4f_{7/2}$ and O $1s$ binding energies (BEs), as well as the lowest intensity from the Au $4f_{7/2}$ signal. Conversely, the oxides prepared at higher temperatures exhibit higher $\Phi$ values and lower BEs for the oxide components. This trend reflects a higher density of metallic species and/or oxygen vacancies acting as interlayer defects, leading to higher screening effects and lower BEs; further annealing promotes the segregation of these impurities, ultimately reverting the system back to a state highly reminiscent of the oxide prepared at lower temperatures. 

 \begin{figure*}[htbp]
	\includegraphics[width=0.8\textwidth]{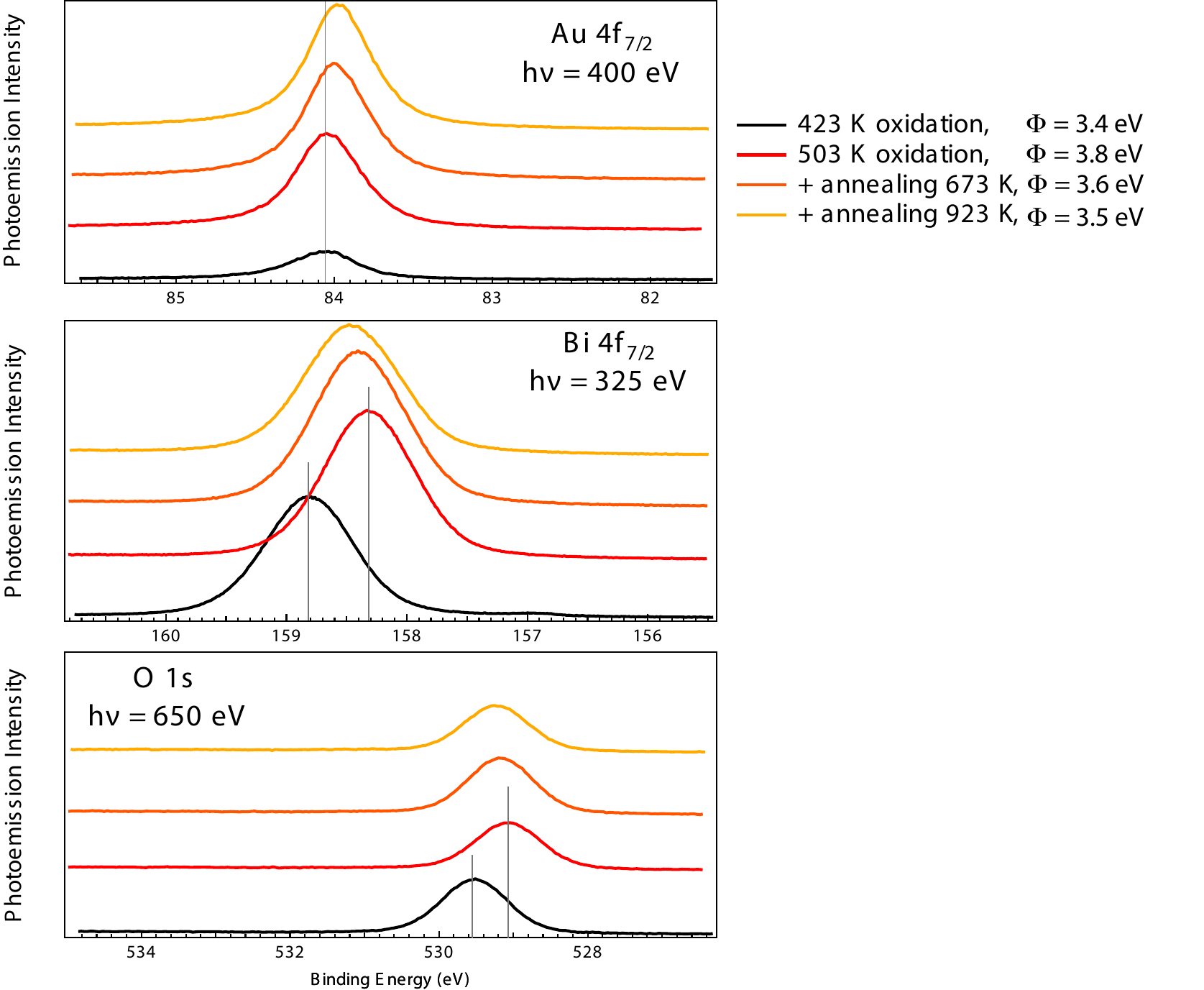}
	\caption{From top to bottom: Au 4$f_{7/2}$, Bi 4$f_{7/2}$ and O 1s  core level spectra acquired at h$\nu=400$~eV,  h$\nu=325$~eV and h$\nu=650$~eV respectively for different oxidation temperature  with the corresponding measured work functions $\Phi$ as explained in the main text. Grey vertical lines are guide to the eye for positioning of the main components.}
	\label{fig:CL_vs_wf}
\end{figure*}

\FloatBarrier





%

\end{document}